\documentclass[12pt,preprint]{aastex}
\usepackage{graphicx}
\usepackage{times}

\def\gsim { \lower .75ex \hbox{$\sim$} \llap{\raise .27ex \hbox{$>$}} }
\def\lsim { \lower .75ex \hbox{$\sim$} \llap{\raise .27ex \hbox{$<$}} }

\newcommand{\plb}{{Phys.~Lett.~B}}
\newcommand{\ijmpd}{{Int.~J.~Mod.~Phys.~D}}
\newcommand{\mpla}{{Mod.~Phys.~Lett.~A}}

\shorttitle{Overcoming Circular Problem for GRB in MCMC}
\shortauthors{Li \emph{et al.}}

\begin{document}

\title{Overcoming the Circular Problem for $\gamma-$ray
Bursts\\
in Cosmological Global Fitting Analysis}

\author{Hong Li\altaffilmark{1}, Jun-Qing Xia\altaffilmark{2}, Jie Liu\altaffilmark{2},
Gong-Bo Zhao\altaffilmark{3}, Zu-Hui Fan\altaffilmark{1} \& Xinmin
Zhang\altaffilmark{2}}

\altaffiltext{1}{Department of Astronomy, School of Physics,
Peking University, Beijing, 100871, P. R. China.}

\altaffiltext{2}{Institute of High Energy Physics, Chinese Academy
of Science, P. O. Box 918-4, Beijing 100049, P. R. China; Email
Address: xiajq@mail.ihep.ac.cn.}

\altaffiltext{3}{Department of Physics, Simon Fraser University,
Burnaby, BC, V5A 1S6, Canada.}

\begin{abstract}
Due to the lack of low redshift long Gamma-Ray Bursts (GRBs), the
circular problem has been a severe obstacle for using GRBs as
cosmological candles. In this paper, we present a new method to
deal with such a problem in Markov Chain Monte Carlo (MCMC) global
fitting analysis. Our methodology is similar to that of
self-calibrations in using clusters of galaxies as cosmological
probes. Assuming that a certain type of correlations between
different observables exists in a subsample of GRBs, for the
parameters involved in the correlation relation, we treat them as
free parameters and determine them simultaneously with
cosmological parameters through MCMC analysis on GRB data together
with other observational data, such as SNe Ia, cosmic microwave
background radiation (CMB) and large-scale structure (LSS). Then
the circular problem is naturally eliminated in this procedure. To
demonstrate the feasibility of our method, we take the Ghirlanda
relation ($E_{\gamma}\propto CE_{\rm peak}^A$) as an example while
keeping in mind the debate about its physical validity. Together
with SNe Ia, WMAP and SDSS data, we include 27 GRBs with the
reported Ghirlanda relation in our study, and perform MCMC global
fitting. We consider the $\Lambda$CDM model and dynamical dark
energy models with equation of state (EoS) $w_{\rm
DE}=w_0+w_1(1-a)$ and the oscillating EoS $w_{\rm DE} = w_{0} +
w_{1}\sin(w_2\ln(a))$, respectively. We also include the curvature
of the universe in our analysis. In each case, in addition to the
constraints on the relevant cosmological parameters, we obtain the
best fit values as well as the distributions of the correlation
parameters $A$ and $C$. We find that the observational data sets
other than GRBs can affect $A$ and $C$ considerably through their
degeneracies with the cosmological parameters. With
CMB+LSS+SNe+GRB data included in the analysis, the results on $A$
and $C$ for different cosmological models are in well agreement
within $1\sigma$ range. The best fit value of $A$ in all models
being analyzed is $A\sim 1.53$ with $\sigma \sim 0.08$. For $C$,
we have the best value in the range of $0.94-0.98$ with
$\sigma\sim 0.1$. It is also noted that the distributions of $A$
and $C$ are generally broader than the priors used in many studies
in literature. Our method can be readily applied to other GRB
relations, which might be better physically motivated.
\end{abstract}



\section{Introduction}
\label{introduction}

Searching for the nature of dark energy has been one of the most
challenging tasks in cosmological studies. Because of the
existence of degeneracies between dark energy parameters and the
other cosmological parameters in different observables,
multi-probe analysis are essential in constraining tightly the
properties of dark energy. In this regard, exploring new probes
has its great importance. On the other hand, thorough
investigations on different probes both observationally and
theoretically are equally important so that we can understand
their validity and limitations in cosmological applications.

GRBs are the most powerful events observed in the cosmos, and can
potentially be used to probe the high-redshift universe. Recently,
several empirical correlations between GRB observables were
reported, which have triggered intensive studies on the
possibility in using GRBs as cosmological known candles (Norris et
al. (2000), Lloyd-Ronning $\&$ Ramirez-Ruiz (2002), Ghirlanda et
al. (2004a), Ghirlanda et al. (2004b), Dai et al. (2004), Xu et
al. (2005), Firmani et al. (2005), Friedman $\&$ Bloom (2005),
Firmani et al. (2006), Schaefer (2007)). Constraints on
cosmological parameters from GRBs alone and in conjunction with
other geometrical probes, including SNe Ia, the shift parameter
from CMB measurements (Wang $\&$ Mukherjee (2006)) and the $A$
parameter for the signals of baryon acoustic oscillation (BAO)
from galaxy redshift surveys (Eisenstein et al. (2005)), have been
analyzed (Su et al. (2006), Wright (2007), Wang et al. (2007)). Li
et al. (2008), for the first time, performed global fitting on the
GRB data with the MCMC technique together with SNe Ia data (Riess
et al. (2007)), and data from WMAP (Spergel et al. (2007)), SDSS
(Tegmark et al. (2004)) and 2dFGRS (Cole et al. (2005)). On the
other hand, the physics behind the empirical correlations are
poorly understood. There are also observational indications that
some of the reported correlations may have potential problems.
Thus there is an ongoing debate for the validity of using GRBs as
cosmological candles. The circular problem has been recognized as
another obstacle in the cosmological applications of GRBs. Up to
now, there have been about $100$ GRBs with measured redshifts, and
few are at low redshifts with known distances. Thus it is lack of
observational data to calibrate, in a cosmology-independent way,
the correlation relations. The reported relations are often given
assuming an input cosmology. Applying such relations to constrain
cosmological parameters leads to the circular problem. Different
methods have been put forward to avoid this problem (Firmani et
al. (2005), Schaefer (2007)). All of them are discussed in the
context of using geometrical constraints only.

In this paper, we present a method to deal with the circular
problem in the MCMC global fitting. It is known that for
constraining cosmological parameters, the most reliable way is to
perform global fitting from observational data directly. Li et al.
(2008) made a first effort to integrate GRBs in the MCMC chains.
However, the GRB data they used are released by Schaefer (2007)
where the distance moduli of GRB samples are not independent of
the input cosmology model and are still subject to the circular
problem. Due to this reason, here we aim at introducing a new
method to get rid of the circular problem of GRBs in order to
avoid biases arising from it, so that the advantage of the MCMC
global fitting can be fully realized. We are aware of the current
debate regarding the cosmological applicability of GRBs (Bloom et
al. (2003), Friedman $\&$ Bloom (2005)). On the other hand, with
both observational and theoretical advances, reliable correlation
relations from sub classes of GRBs may eventually emerge. With
these considerations in mind, we have made our method as general
as possible. It is not limited to any specific correlations.
However, to demonstrate its feasibility, we have to work on a
concrete example. We choose the Ghirlanda relation (Firmani et al.
(2006)) in our present study. Our method can be readily applied to
other correlations. The cosmological models to be analyzed include
the $\Lambda$CDM model and dynamical dark energy models with the
EoS following $w_{\rm DE}=w_0+w_1(1-a)$ and $w_{\rm DE}= w_{0} +
w_{1}\sin(w_2\ln(a))$, respectively. The dark energy perturbations
are fully taken into account.


\section{Methodology}

For the paper to be self-contained, in this section, we firstly
describe briefly the Ghirlanda relation, and then we present our
analyzing method dealing with the circular problem of GRBs in the
MCMC fitting procedure. The general global fitting procedure will
be explained in the second part of the section.

\subsection{The Ghirlanda relation and the method}

Keeping in mind its general applicability, we have to choose a
specific correlation for GRBs as an example to show quantitatively
the feasibility of our method in the MCMC global fitting analysis.
Among the reported correlations, the Ghirlanda relation is the one
that has been used most extensively in constraining cosmology due
to its relatively small scatters and the relatively large number
of data points available. Recently, there have been intensive
arguments questioning this relation largely because of the
observed complexities of the X-ray light curves, which lead to the
difficulties in identifying jet break features. On the other hand,
it has been pointed out that the X-ray and the optical emissions
of GRB afterglows may have different origins, and thus can behave
differently. In the recent study of Ghirlanda et al. (2007), they
emphasize that the jet break features should be considered only if
they appear in optical light curves. With the awareness of these
debates, we here adopt the Ghirlanda relation in our analysis, and
focus on the method for the circular problem, instead of on the
cosmological constraints from GRB data.

The Ghirlanda relation, or the $E_{\rm peak}-E_{\gamma}$
correlation, relies on the jet break feature to calculate the jet
opening angle $\theta_{\rm jet}$, which, in turn, is crucial in
correcting the GRB prompt emission energy for the collimation
effects. Here we adopt the homogeneous medium model with
\begin{equation}
\theta_{\rm jet} = 0.161\left( {\frac{{t_{\rm j,d} }}{{1 + z}}}
\right)^{3/8} \left( {\frac{{n_0 }{\eta _\gamma }}{{E_{{\rm
iso},52} }}} \right)^{1/8}~, \label{theta}
\end{equation}
where $z$ is the redshift, $\eta _\gamma$ is the radiative
efficiency, $t_{{\rm j},d}$ is the break time in days with
$t_{{\rm j,d}}=t_{\rm j}/1\,{\rm day}$, $n_0$ is the number
density of particles in the surrounding interstellar medium with
$n_0=n/1\,{\rm cm}^{-3}$ and $E_{{\rm iso}}$ is the
isotropic-equivalent energy of GRBs with $E_{{\rm iso},52}=E_{{\rm
iso}}/10^{52}{\rm ergs}$. The quantity $E_{{\rm iso}}$ is related
to the observed fluence $S_\gamma$ in units of erg/cm$^{2}$ as
follows:
\begin{equation}
E_{{\rm iso}}=\frac{4\pi d_L^2S_\gamma k}{1+z}~, \label{eiso}
\end{equation}
where $d_L$ is the luminosity distance at redshift $z$, and $k$ is
a multiplicative correction related the observed bandpass to a
standard rest-frame bandpass (1-$10^4$ keV in this paper) (Bloom
et al. (2001)). The collimation corrected energy $E_{\gamma}$ is
\begin{equation}
E_{\gamma}  = (1 - \cos \theta_{\rm jet} )E_{\rm iso}~.
\label{er}\end{equation}

Following Xu et al. (2005), we write the $E_{\rm peak}-E_{\gamma}$
correlation in the following form:
\begin{equation}
\frac{E_{\gamma}}{10^{50}{\rm ergs}}=C\left(\frac{E_{\rm
peak}}{100\,{\rm keV}}\right)^{A}, \label{relation}
\end{equation}
where the parameters $A$ and $C$ are assumed to be constant, and
$E_{\rm peak}=E^{\rm obs}_{\rm peak}(1+z)$. It is seen that
besides the direct observables, the luminosity distance $d_L$
comes in through Eqs.(1-2). If, for a set of GRBs, their distances
can be determined independently, one can calibrate the correlation
relation (4) and find the values of $A$ and $C$ from observations
directly. Then the cosmology-independent correlation can be used
to estimate $d_L$ for other GRBs and further to constrain
cosmology. Unfortunately we lack of GRBs with known distances.
Therefore in order to obtain values of $A$ and $C$, one has to
assume a cosmological model to calculate $d_L$. The circular
problem arises when the luminosity distances derived from such a
cosmology-dependent correlation relation are used as cosmological
candles. Different methods have been discussed to avoid the
circular problem in grid-based $\chi^2$ analysis involving only
geometrical probes. In the following we describe our method to
deal with the circular problem in MCMC global fitting procedures.

The essential of our method is that for an assumed functional form
of a correlation relation such as in Eq.(4), we set the
correlation parameters free in our analyzing process instead of
using the values reported in other studies. Simultaneously with
the cosmological parameters, their values are determined through
global fittings with GRB data and other data sets, including CMB,
LSS and SNe. Specifically, for each element on MCMC chains with a
set of parameters $(x_i, A, C)$, where $x_i$, $i=1\cdots n$ are
cosmological parameters we are interested in, the ``observed"
luminosity distance for each GRB is obtained through Eqs.(1-4) as:
\begin{equation}
d_L  = 7.575~\frac{{(1 + z)C^{2/3} [E_{\rm peak}^{\rm obs} (1 +
z)/100\,{\rm keV}]^{2A/3} }}{{(kS_\gamma t_{\rm j,d} )^{1/2} (n_0
\eta _\gamma )^{1/6} }}~{\rm{Mpc}}~, \label{dl}
\end{equation}
where the small angle approximation with $\theta_{\rm jet}\ll 1$
has been applied. With the assumption that all the GRB observables
are independent of each other with gaussian distributed errors,
the uncertainty for each of these ``data" points is estimated as
follows (Friedman $\&$ Bloom (2005))
\begin{eqnarray}
 \left( {\frac{{\sigma _{d_L } }}{{d_L }}} \right)^2   &=&  \frac{1}{4}\left[
 {\left( {\frac{{\sigma _{S_\gamma} }}{{S_\gamma}}} \right)^2  + \left( {\frac{{\sigma _k
  }}{k}} \right)^2 } \right] +\frac{1}{4}\frac{{1 }}{{(1 - \sqrt
   {C_{\theta}})^2 }}\left( {A\frac{{\sigma _{E_{\rm peak}^{\rm obs} } }}{{E_{\rm peak}^{\rm obs} }}} \right)^2 \nonumber\\ &+& \frac{1}{4}\frac{C_{\theta}}{{(1 - \sqrt
   {C_{\theta}})^2 }} \left[ {\left( {\frac{{3\sigma _{t_{\rm j,d}} }}{{t_{\rm j,d}}}} \right)^2
    + \left( {\frac{{\sigma _{n_0} }}{{n_0}}} \right)^2}+ \left( {\frac{{\sigma _{\eta _\gamma}
}}{{\eta _\gamma}}} \right)^2 \right]~\label{error},
\end{eqnarray}
where
\begin{equation}
C_\theta   = \left[ {\theta \sin \theta /(8 - 8\cos \theta )}
\right]^2.
\end{equation}We take $\eta_\gamma=0.2$ and
$\sigma_{\eta_\gamma}=0$ throughout this paper (Frail et al.
(2001)). It is noted that when Eq.(6) is used to calculate
$\sigma_{d_L}$, it is implicitly assumed that the correlation
Eq.(4) has no additional scatters besides the uncertainties for
the parameters $A$ and $C$. Considering the distance modulus, we
have
\begin{equation}
\mu_{\rm obs}=5\log_{10} d_L+25~,
\end{equation}
and
\begin{equation}
\sigma_{\mu_{\rm
obs}}=\frac{5}{\ln10}\left(\frac{\sigma_{d_L}}{d_L}\right)~.
\end{equation}

In order to constrain the cosmological parameters $x_{\rm i}$, we
have marginalized the free parameters $A$ and $C$, and finally we
get the probability for a certain cosmological parameter $x_{\rm
i}$ as follows:
\begin{equation}
P(x_i) = \int P(x_i|x_j\cdots A,C)P(x_j)\cdots P(A)P(C)dx_j\cdots
dAdC~,
\end{equation}
which are related to $\chi^2$ given by the statistic results of
the observational data via $P\propto e^{-\chi^2/2}$. And the
$\chi^2$ contributed by GRB ``data" at the point $(x_{\rm i}, A,
C)$ is then computed as
\begin{equation}
\chi^2(x_{\rm i},A,C)=\sum\left[\frac{\mu_{\rm th}(x_{\rm
i})-\mu_{\rm obs}(x_{\rm i},A,C)}{\sigma_{\mu_{\rm
obs}}}\right]^2~,
\end{equation}
where the summation is over the number of GRB data points.

We use $27$ GRBs, which are reportedly to satisfy the $E_{\rm
peak}-E_{\gamma}$ relation, in our study. The relevant data are
listed in Table 1. In Table 1, the data are mostly from Ghirlanda
et al. (2007) except for GRB050505 and GRB060210. For these two
GRBs, we take the data from Schaefer
(2007)\footnote{Communications with Ghirlanda, 2007.}.

\begin{table*}{\footnotesize

\caption{Sample of 27 GRBs}
\begin{center}

\begin{tabular}{ccccccc}

\hline\hline

&&~~$E^{\rm obs}_{\rm peak}$($\sigma_{E^{\rm obs}_{\rm
peak}}$)~~&$S_{\gamma}$($\sigma_{S_{\gamma}}$)&~~~$t_{\rm
j,d}$($\sigma_{t_{\rm j,d}}$)~~~
&${}^a$$n$($\sigma_n$)&~~${}^b$Reference~~~\\

~~~~GRB~~~~ & $z$ & (keV) & ~~($10^{-6}$ergs/$cm^2$)~~ & (days)&
($cm^{-3}$)& ($z,E^{\rm obs}_{\rm peak},S_{\gamma},t_{\rm
j,d},n$) \\

\hline

$970828$ & $0.9578$ & $297.7~[59.5]$ & $96.0~[9.6]$ & $2.2~[0.4]$
& $3.0~[2.4]$ &01,28,41,01,no\\
$980703$ & $0.966$ & $254~[50.8]$ & $22.6~[2.3]$ & $3.4~[0.5]$ &
$28.0~[10.0]$ &02,28,41,48,48\\
$990123$ & $1.600$ & $780.8~[61.9]$ & $300~[40]$ & $2.04~[0.46]$ &
$3.0~[2.4]$ &03,29,29,49,no\\
$990510$ & $1.619$ & $161.5~[16.1]$ & $19~[2]$ & $1.6~[0.2]$ &
$0.29~[0.14]$ &04,29,29,50,68\\
$990705$ & $0.8424$ & $188.8~[15.2]$ & $75~[8]$ & $1.0~[0.2]$ &
$3.0~[2.4]$ &05,29,29,51,no\\
$990712$ & $0.4331$ & $65~[11]$ & $6.5~[0.3]$ & $1.6~[0.2]$ &
$3.0~[2.4]$ &06,29,29,52,no\\
$991216$ & $1.020$ & $317.3~[63.4]$ & $194~[19]$ & $1.2~[0.4]$ &
$4.7~[2.8]$ &07,28,41,53,68\\
$010222$ & $1.480$ & $309~[12]$ & $93~[3]$ & $0.93~[0.1]$ &
$3.0~[2.4]$ &08,29,42,54,no\\
$011211$ & $2.140$ & $59.2~[7.6]$ & $5.0~[0.5]$ & $1.56~[0.02]$ &
$3.0~[2.4]$ &09,30,41,55,no\\
$020124$ & $3.200$ & $120.0~[22.6]$ & $6.8~[0.68]$ & $3.0~[0.4]$ &
$3.0~[2.4]$ &10,31,31,56,no\\
$020405$ & $0.690$ & $192.5~[53.8]$ & $74~[0.7]$ & $1.67~[0.52]$ &
$3.0~[2.4]$ &11,32,32,32,no\\
$020813$ & $1.255$ & $142~[13]$ & $97.9~[10]$ & $0.43~[0.06]$ &
$3.0~[2.4]$ &12,31,31,57,no\\
$021004$ & $2.332$ & $79.8~[30]$ & $2.6~[0.6]$ & $4.74~[0.14]$ &
$30.0~[27.0]$ &13,33,33,58,69\\
$021211$ & $1.006$ & $46.8~[5.5]$ & $3.5~[0.1]$ & $1.4~[0.5]$ &
$3.0~[2.4]$ &14,34,34,59,no\\
$030226$ & $1.986$ & $97~[20]$ & $5.61~[0.65]$ & $1.04~[0.12]$ &
$3.0~[2.4]$ &15,33,33,60,no\\
$030328$ & $1.520$ & $130.2~[13.9]$ & $37~[1.4]$ & $0.8~[0.1]$ &
$3.0~[2.4]$ &16,33,33,61,no\\
$030329$ & $0.1685$ & $67.9~[2.2]$ & $163~[10]$ & $0.5~[0.1]$ &
$1.0~[0.11]$ &17,35,35,62,70\\
$030429$ & $2.6564$ & $35~[9]$ & $0.85~[0.14]$ & $1.77~[1]$ &
$3.0~[2.4]$ &18,33,33,18,no\\
$041006$ & $0.716$ & $63.4~[12.7]$ & $19.9~[1.99]$ & $0.16~[0.04]$
&$3.0~[2.4]$ &19,36,43,63,no\\
$050401$ & $2.900$ & $128.5~[30]$ & $19.3~[0.4]$ & $1.5~[0.5]$
&$3.0~[2.4]$ &20,37,40,27,no\\
$050416$ & $0.653$ & $17.3~[5]$ & $0.35~[0.03]$ & $1.0~[0.7]$
&$3.0~[2.4]$ &21,38,40,27,no\\
$050505$ & $4.270$ & $70~[{}^{+140}_{-24}]$ & $4.1~[0.4]$ &
$0.21~[0.04]$&$3.0~[2.4]$ &22,39,44,64,no\\
$050525$ & $0.606$ & $79~[3.5]$ & $20.1~[0.5]$ & $0.28~[0.12]$
&$3.0~[2.4]$ &23,37,45,65,no\\
$050820$ & $2.612$ & $246~[{}^{+76}_{-40}]$ & $52.7~[6.9]$ &
$15.2~[8]$&$3.0~[2.4]$ &24,37,40,27,no\\
$060210$ &$3.910$ & $149~[{}^{+400}_{-35}]$ & $7.7~[0.4]$ &
$0.33~[0.08]$&$3.0~[2.4]$ &25,40,46,66,no\\
$060526$ & $3.210$ & $24.94~[5]$ & $0.49~[0.06]$ & $2.77~[0.3]$
&$3.0~[2.4]$ &26,40,47,67,no\\
$060614$ & $0.125$ & $48.9~[40]$ & $21.7~[0.4]$ & $1.38~[0.04]$
&$3.0~[2.4]$ &27,27,27,27,no\\

\hline\hline

\end{tabular}
\end{center}}

\tiny

{${}^a$The circumburst densities and errors from broadband
modelling of the afterglow light curves. If no available
the value of $n$ is taken as $3.0\pm2.4$ cm$^{-3}$.\\
${}^b$References for the GRBs data in the table: (01)Djorgovski,
S. G. et al. 2001, ApJ, 562, 654; (02)Djorgovski, S.G. et al.
1998, ApJ, 508, L17; (03)Kulkarni, S. R. et al. 1999, Nature, 398,
389;
  (04)Vreeswijk, P. M. et al. 2001, ApJ, 546, 672;
   (05)Le Floc'h, E. 2002, ApJ, 581, L81;
    (06)Vreeswijk, P. M. et al. 2001, ApJ, 546, 672;
     (07)Piro, L. et al. 2000, Science, 290, 955;
      (08) Fruchter, A. et al. 2001a, GCN 1029;
      (09)Holland, et al. 2002, AJ, 124, 639;
(10)Hjorth, J. et al. 2003, ApJ, 597, 699;
 (11)Price, P. A. et al. 2003a, ApJ, 589, 838;
  (12)Barth, A. J. et al. 2003, ApJ, 584, L47;
   (13)Matheson, T. et al. 2003, ApJ, 582, L5;
    (14)Vreeswijk, P. M. et al. 2003, GCN, 1785;
     (15)Greiner, J. et al. 2003a, GCN, 1886;
     (16)Rol, E. et al. 2003, GCN, 1981;
      (17)Greiner, J. et al. 2003b, GCN, 2020;
       (18)Jakobsson, P. et al. 2004, A\&A, 427, 785;
        (19)Fugazza, D. et al. 2004, GCN 2782;
         (20) Fynbo, J. P. U. et al. 2005b, GCN 3176;
         (21) Cenko, S. B. et al. 2005, GCN 3542;
         (22)Berger, E. et al. 2005b, GCN 3368;
          (23) Foley, R. J. et al. 2005, GCN 3483;
          (24)Prochaska, J. X. et al. 2005b, GCN 3833;
           (25)Cucchiara, A. et al. 2006a, GCN 4729;
            (26)Berger. E. \& Gladders, M. 2006, GCN 5170;
             (27)G. Ghirlanda, L. Nava, G. Ghisellini and C.
 Firmani, astro-ph/0702352;
  (28)Jimenez, R., Band, D. L., \& Piran, T., 2001, ApJ, 561, 171;
(29)Amati, L. et al. 2002, A\&A, 390, 81; (30)Amati, L. 2004,
astro-ph/0405318;
 (31)Barraud, C. et al. 2003, A\&A, 400, 1021;
  (32)Price, P. A. et al. 2003a, ApJ, 589, 838;
   (33)Sakamoto, T. et al., 2004b, astro-ph/0409128;
    (34)Crew, G. B. et al. 2003, ApJ, 599, 387;
     (35)Vanderspek, R. et al. 2004, AJ, 617, 1251;
      (36)HETE 2006, http://space.mit.edu/HETE/Bursts/;
       (37)Krimm, H. et al. 2006a, in Gamma-Ray Bursts
     in the Swift Era, eds S. S. Holt, N. Gehrels, and J. A. Nousek (AIP Conf. Proc. 836), pp. 145-148;
      (38)Sakamoto, T. et al. 2006b, ApJ, 636, L73;
       (39)Krimm, H. et al. 2005, GCN 3134;
        (40)Schaefer, astro-ph/0612285;
(41)Bloom, J. S., Frail, D. A., \& Kulkarni, S. R. 2003, ApJ, 594,
674;
 (42)Frontera, F. et al. 2001, GCN 1215;
  (43)Galassi, M. et al. 2004, GCN 2770;
   (44)Hullinger, D. et al. 2005, GCN 3364;
    (45)Cummings, J. et al. 2005, GCN 3479;
     (46)Sakamoto, T. et al. 2006d, GCN 4748;
      (47)Markwardt, C. et al. 2006, GCN 5174;
       (48)Frail, D. A. et al. 2003, ApJ, 590, 992;
        (49)Kulkarni, S. R. et al. 1999, Nature, 398, 389;
         (50)Stanek, K. Z. et al. 1999, ApJ, 522, L39;
          (51)Masetti, N. et al. 2000, A\&A, 354, 473;
           (52)Bjornsson G. et al. 2001, ApJ, 552, L121;
            (53)Halpern, J. P., et al. 2000, ApJ, 543, 697;
(54)Jakobsson, P. et al. 2003, A\&A, 408, 941;
 (55)Jakobsson, P. et al. 2003, A\&A, 408, 941;
  (56)Berger, E. et al. 2002, ApJ, 581, 981;
   (57)Barth, A. J. et al. 2003, ApJ, 584, L47;
    (58)Holland, et al. 2003, AJ, 125, 2291;
(59)Holland, et al. 2004, astro-ph/0405062;
 (60)Klose, S. et al. 2004, AJ, 128, 1942;
  (61)Andersen, M. I. et al. 2003, GCN, 1993;
   (62)Price, P. A. et al. 2003b, Nature, 423, 844;
    (63)Stanek, K. Z. et al. 2005, ApJ, 626, L5;
     (64)Hurkett, C. P. et al. 2006, MNRAS, 368, 1101;
(65)Blustin, A. J. et al. 2006, ApJ, 637, 901;
 (66)Dai, X. \& Stanek, Z. 2006, GCN 5147;
  (67)Moretti, A. et al. 2006, GCN 5194;
   (68)Panaitescu, A., \& Kumar, P. 2002, ApJ, 571, 779;
    (69)Schaefer, B. E. 2003, ApJ, 583, L67 (S03);
(70)Tiengo, A. et al. 2003, A\&A, 409, 938.\\}
\end{table*}


\subsection{Global fitting program}

Different observations play complementary roles in the
determination of cosmological parameters. Their combination can
effectively break out the degeneracies between different
parameters, and therefore can deliver much better constraints on
cosmology than any single probe can. For different observable, it
is important to understand the main factors that affect the
determination of interested parameters. The information on these
elements extracted from observational data is very useful. Under
certain conditions, the extracted values of these factors can be
used to probe cosmology without invoking complicated observational
data, which could greatly simplify the analyzing procedures. The
two important examples are the shift parameter from CMB
observations and the BAO $A$ parameter from galaxy redshift
surveys, and they have been widely used in constraining properties
of dark energy. On the other hand, however, careful attentions
must be paid to the conditions under which the extracted
information is obtained. Inappropriate using of these pieces of
information can lead to biased conclusions on the values of
cosmological parameters. Therefore the most reliable way in
determining cosmology is to perform global fitting analysis using
observational data directly. Our global fitting analysis are based
on the publicly available MCMC package CosmoMC (Lewis $\&$ Bridle
(2002)). We have made modifications according to our own research
purposes. Besides the modifications described in part A of this
section, which are specific for the circular problem of GRBs, we
include dark energy perturbations in our general analyzing
program.

For dark energy models with equation of state $w\neq -1$, the
perturbations inevitably exist. While the effects of dark energy
perturbations are yet to be fully explored, it is generally
believed that they may only show their influences at near-horizon
scales. For CMB anisotropy, the power spectrum at low $l$ (large
angular scale) is affected by the late Integrated Sachs-Wolfe
Effect (ISW), whose strength depends on the properties of dark
energy in a flat universe. Thus large-scale anisotropy is
important and the perturbations can play roles in dark energy
studies. It is well recognized that in the fluid approach, there
is a divergence problem at $w=-1$ when dark energy perturbations
are included. On the other hand, there are observational
indications that the equation of state of dark energy may cross
$-1$ during the evolutionary history of the universe (Huterer $\&$
Cooray (2005), Feng et al. (2005), Xia et al. (2006), Zhao et al.
(2007a)). Thus for dark energy perturbations, the divergence
problem must be carefully dealt with. From our analysis on
two-field quintom models (Feng et al. (2005), Zhang et al.
(2006)), in which $w$ crosses $-1$ can be realized, we find that
the dark energy perturbations are well behaved at $w=-1$. Thus the
divergence in the fluid treatment should be a mathematical one
instead of a physical one. Along this line of thinking, we develop
a scheme in the fluid approach to avoid the divergence problem
(Zhao et al. (2005), Xia et al. (2006)).

In the conformal Newtonian gauge, the perturbation equations of
dark energy are:
\begin{eqnarray}
    \dot\delta&=&-(1+w)(\theta-3\dot{\Phi})
    -3\mathcal{H}(c_{s}^2-w)\delta~~, \label{dotdelta}\\
\dot\theta&=&-\mathcal{H}(1-3w)\theta-\frac{\dot{w}}{1+w}\theta
    +k^{2}(\frac{c_{s}^2\delta}{{1+w}}+ \Psi)~~ . \label{dottheta}
\end{eqnarray}
where $\delta$ and $\theta$ are the energy density and
velocity perturbations, respectively. The divergence at $w=-1$ can
be seen from the second equation. To handle this problem, we
introduce a small constant $\epsilon$, and divid $w$ into three
parts with 1) $ w > -1 + \epsilon$; 2) $-1 + \epsilon \geq w
\geq-1 - \epsilon$; and 3) $w < -1 -\epsilon $, respectively. For
regions 1) and 3), the perturbations are analyzed following the
equations. For region 2), we need a special treatment. We match
the perturbation quantities of region 2) to regions 1) and 3) at
the corresponding boundaries, and set
\begin{equation}\label{dotx}
  \dot{\delta}=0 ~~,~~\dot{\theta}=0 .
\end{equation}
within the region. Thus there are discontinuities in the
derivatives in region 2). But with small enough $\epsilon$, the
discontinuities have negligible effects. We compare the results
from this analysis with those of two-field quintom models and find
that with $\epsilon\le 10^{-5}$, the perturbations of the quintom
models can well be reproduced by this fluid approach. For more
details of this method we refer the readers to our previous
companion papers (Zhao et al. (2005), Xia et al. (2006)). Thus we
set $\epsilon= 10^{-5}$ in our studies.

We consider three cosmological models, the $\Lambda$CDM model
including the curvature term, and dynamical dark energy models
with the equation of state parameterized respectively as
\begin{eqnarray}
{\rm I})&~&~~~~w_{\rm DE}(a) = w_{0} + w_{1}(1-a)~,\\
{\rm II})&~&~~~~w_{\rm DE}(a) = w_{0} + w_{1}\sin(w_2\ln(a))~,
\end{eqnarray}
where $a=1/(1+z)$ is the scale factor and $w_{1}$ characterizes
the ``running" of the EoS. For the parametrization I (Para I), we
include the curvature term. In the parametrization II (Para II),
we are limited in the flat universe and fix $w_2=3\pi/2$ for not
introducing too many parameters during the fitting process.

Our most general parameter space is then:
\begin{equation}
\label{parameter} {\bf P} \equiv (\omega_{b}, \omega_{c},
\Omega_K, \Theta_{s}, \tau,  w_{0}, w_{1}, n_{s},
\ln(10^{10}A_{s}), A, C)
\end{equation}
where $\omega_{b}\equiv\Omega_{b}h^{2}$,
$\omega_{c}\equiv\Omega_{c}h^{2}$, $\Omega_K$ represents the
contribution of the curvature term to the total energy budget,
$\Theta_{s}$ is the ratio (multiplied by 100) of the sound horizon
at decoupling to the angular diameter distance to the last
scattering surface, $\tau$ is the optical depth due to
re-ionization, $w_0$ and $w_1$ is the parameters of the EoS of
Dark Energy, $A_{s}$ and $n_{s}$ characterize the power spectrum
of primordial scalar perturbations, $A$ and $C$ are the free
parameters related to the $E_{\rm peak}-E_{\gamma}$ correlation.
For the $\Lambda$CDM models, $w_0=-1, w_1=0$.

We vary the above parameters and fit to the observational data
with the MCMC method. For the pivot of the primordial spectrum we
set $k_{s0}=0.05$Mpc$^{-1}$. The following weak priors are taken:
$\tau< 0.8$, $0.5<n_s<1.5$, $-3<w_0<3$, $-5<w_1<5$, $0.5<A<2.5$
and $0.01<C<2.5$. We impose a tophat prior on the cosmic age as 10
Gyr $< t_0 <$ 20 Gyr. Furthermore, we make use of the Hubble Space
Telescope (HST) measurement of the Hubble parameter $H_{0}\equiv
100$h~km~s$^{-1}$~Mpc$^{-1}$ by multiplying the likelihood by a
Gaussian likelihood function centered around $h=0.72$ with a
standard deviation $\sigma = 0.08$ (Freedman et al. (2001)). We
also adopt a Gaussian prior on the baryon density
$\Omega_{b}h^{2}=0.022\pm0.002$ ($1\sigma$) from Big Bang
Nucleosynthesis (Burles et al. (2001)).

In our calculations, we take the total likelihood to be the
products of the separate likelihoods (${\bf \cal{L}}_i$) of CMB,
LSS, SNIa and GRBs. For CMB, we include the three-year WMAP
(WMAP3) data and compute the likelihood with the routine supplied
by the WMAP team (Spergel et al. (2007)). For the Large Scale
Structure information, we have used the Sloan Digital Sky Survey
(SDSS) luminous red galaxy (LRG) sample (Tegmark et al. (2006)).
To minimize the nonlinear effects, we have only used the first
$15$ bins, $0.0120<k_{\rm eff}<0.0998$, which are supposed to be
well within the linear regime. For SNe Ia, we mainly present the
results with the recently released Essence 192 sample supernovae
published in Miknaitis et al. (2007) and Davis et al. (2007). In
the calculation of the likelihood from SNe Ia data, we marginalize
over the nuisance parameter (Goliath et al. (2001), Di Pietro $\&$
Claeskens (2003)).

For each regular calculation, we run 8 independent chains
comprising of $150,000-300,000$ chain elements, and spend
thousands of CPU hours on a supercomputer. The average acceptance
rate is about $40\%$. We test the convergence of the chains by
Gelman and Rubin criteria (Gelman $\&$ Rubin (1992)) and find that
$R-1$ is on the order of $0.01$, which is much more conservative
than the recommended value $R-1<0.1$.


\begin{table*}
\label{table1}

TABLE II. Constraints on the EoS of dark energy and some
background parameters from the current observations with and
without GRBs. Note that Para I and Para II represent $w_{\rm
DE}(a) = w_{0} + w_{1}(1-a)$ and $w_{\rm DE}(a) = w_{0} +
w_{1}\sin(3\pi/2\ln(a))$ respectively. For the current constraints
we have shown the mean values $1\sigma$ (Mean).

\tiny
\begin{center}
\begin{tabular}{|c|cc|cc|cc|cc|}
\hline

&\multicolumn{2}{c|}{$\Lambda$CDM} &\multicolumn{2}{c|}{~Para~I~}
&\multicolumn{2}{c|}{~Para~I~+$\Omega_k$}&\multicolumn{2}{c|}{~Para~II~} \\

\hline

&\multicolumn{1}{c|}{GRB only}&\multicolumn{1}{c|}{ combined all}&\multicolumn{1}{c|}{with GRB}&\multicolumn{1}{c|}{without GRB}&\multicolumn{1}{c|}{with GRB}&\multicolumn{1}{c|}{without GRB}&\multicolumn{1}{c|}{with GRB}&\multicolumn{1}{c|}{without GRB}\\

\hline

$w_0$ & $-1$ & $-1$ & $-1.04^{+0.17}_{-0.16}$ & $-1.01^{+0.18}_{-0.17}$ & $-1.05\pm0.17$ & $-1.02\pm0.17$ & $-0.937^{+0.137}_{-0.147}$ & $-0.968^{+0.148}_{-0.154}$\\

\hline

$w_1$ & $0$ & $0$ & $0.344^{+0.641}_{-0.664}$ & $0.192^{+0.741}_{-0.729}$ & $0.273^{+0.723}_{-0.758}$ & $0.228^{+0.728}_{-0.771}$ & $0.016^{+0.187}_{-0.207}$ & $-0.025^{+0.212}_{-0.221}$\\

\hline

$~\Omega_{\rm DE}~$ & $0.748^{+0.248}_{-0.748}$ & $0.761\pm0.017$ & $0.760\pm0.020$ & $0.762\pm0.020$ & $0.755\pm0.023$ & $0.758\pm0.022$ & $0.764^{+0.021}_{-0.020}$ & $0.764^{+0.022}_{-0.021}$\\

\hline

$\Omega_k$ & $-0.235^{+0.374}_{-0.347}$ & $0$ & $0$ & $0$ & $-0.002\pm0.014$ & $0.001\pm0.014$ & $0$ & $0$\\

\hline

$A$ & $1.51\pm0.0836$ & $1.54^{+0.0782}_{-0.0768}$ & $1.53^{+0.0781}_{-0.0758}$ & $0$ & $1.53^{+0.0771}_{-0.0769}$ & $0$ & $1.54^{+0.0774}_{-0.0736}$ & $0$\\

\hline

$C$ & $0.912^{+0.408}_{-0.352}$ & $0.943^{+0.106}_{-0.108}$ & $0.963\pm0.109$ & $0$ & $0.983^{+0.134}_{-0.139}$ & $0$ & $0.939^{+0.102}_{-0.0996}$ & $0$\\

\hline
\end{tabular}
\end{center}

\end{table*}

\begin{figure}[htbp]
\begin{center}
\includegraphics[scale=0.8]{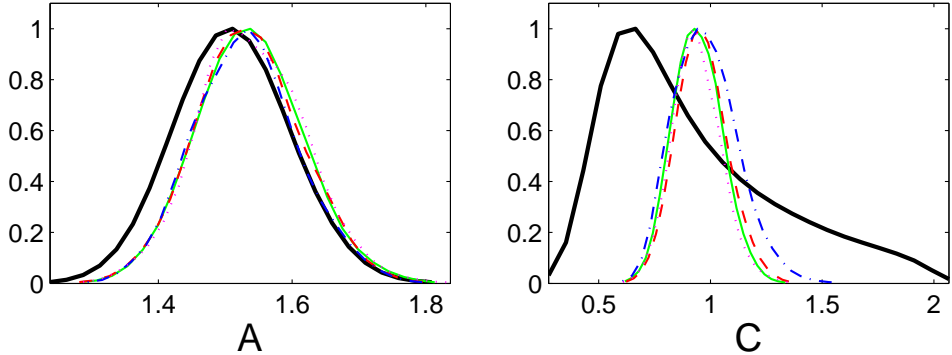}
\caption{1-d posterior constraints for the parameters $A$ and $C$
obtained via MCMC methods. The black solid lines are for the
non-flat $\Lambda$CDM model using GRBs data only. The colored
lines are the results from joint analysis of CMB+LSS+SNe+GRB. The
green solid, red dashed, blue dash-dotted, and magenta dotted
lines are for the $\Lambda$CDM, flat model with dark energy Para
I, non-flat model with dark energy Para I, and flat model with
dark energy Para II, respectively. \label{fig1}}
\end{center}
\end{figure}

\section{Results and Discussions}

In this section, we present our main results. Firstly we consider
the constraint on the non-flat $\Lambda$CDM model using GRB data
only. In order to study the dependence of $A$ and $C$ on the
cosmology models, we combine the GRB data with other cosmological
observational data, such as CMB, LSS and SN, to constrain
different dark energy models, namely the $\Lambda$CDM model and
the dynamical dark energy models, Para I and Para II. Considering
the degeneracy between the dark energy parameters and the
curvature of universe (Zhao et al. (2007b), Clarkson et al.
(2007)), we also consider the non-flat case in the dynamical dark
energy model Para I. During these calculations, we focus on
discussions of constraints on $A$ and $C$, and the effects of GRB
data on the determination of cosmological parameters.

In Figure 1, we show the one-point likelihood function for $A$ and
$C$, respectively. The black solid line in each panel represents
the result for the non-flat $\Lambda$CDM model using GRB data
only. The colored lines are the results from combined analysis of
CMB+LSS+SNe+GRB for different cosmological models. The green
solid, red dashed, blue dash-dotted, and magenta dotted lines are
for the $\Lambda$CDM, flat model with dark energy parametrization
I, non-flat model with dark energy Para I, and flat model with
dark energy Para II, respectively.

\begin{figure}[htbp]
\begin{center}
\includegraphics[scale=0.4]{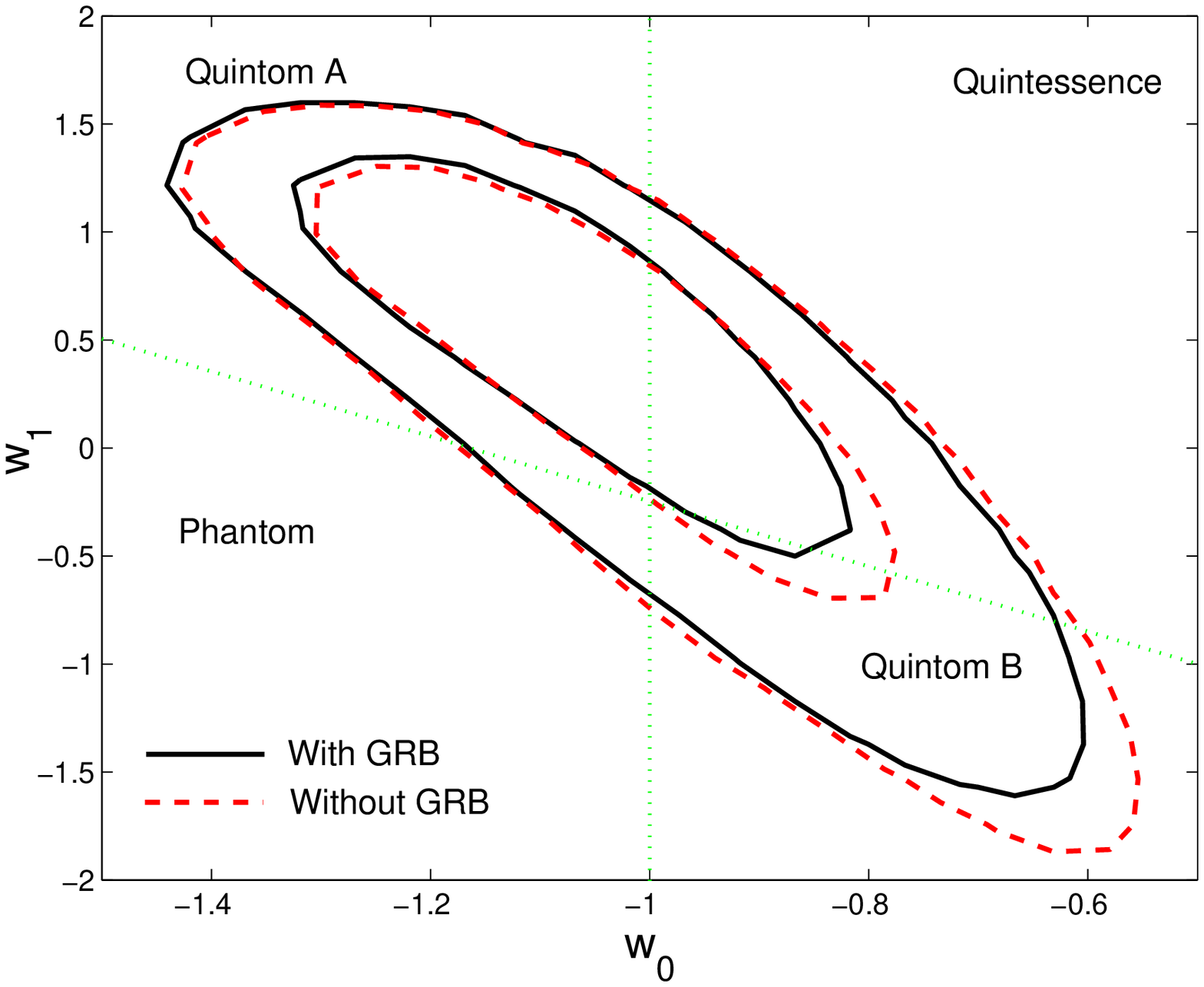}
\includegraphics[scale=0.4]{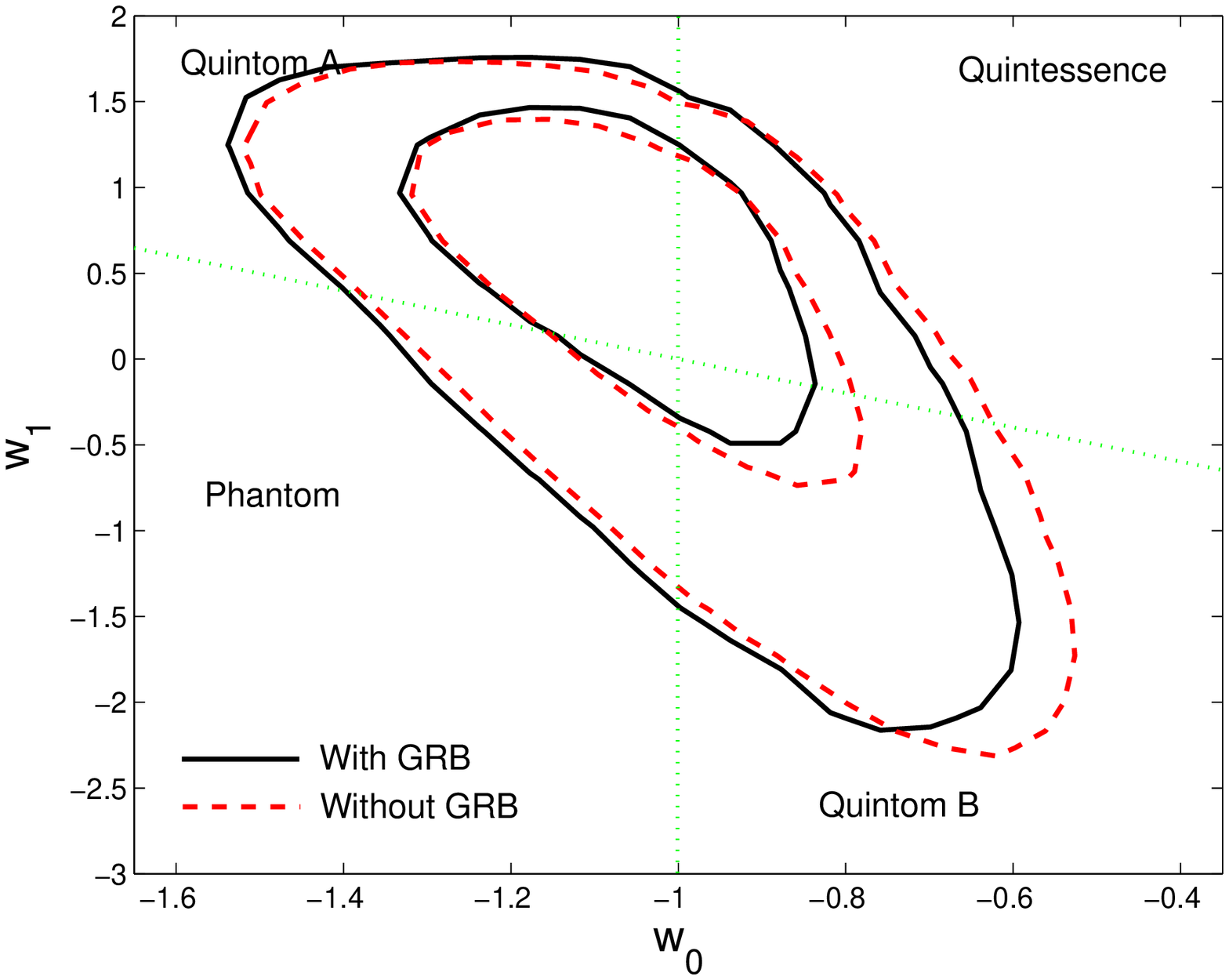}
\includegraphics[scale=0.4]{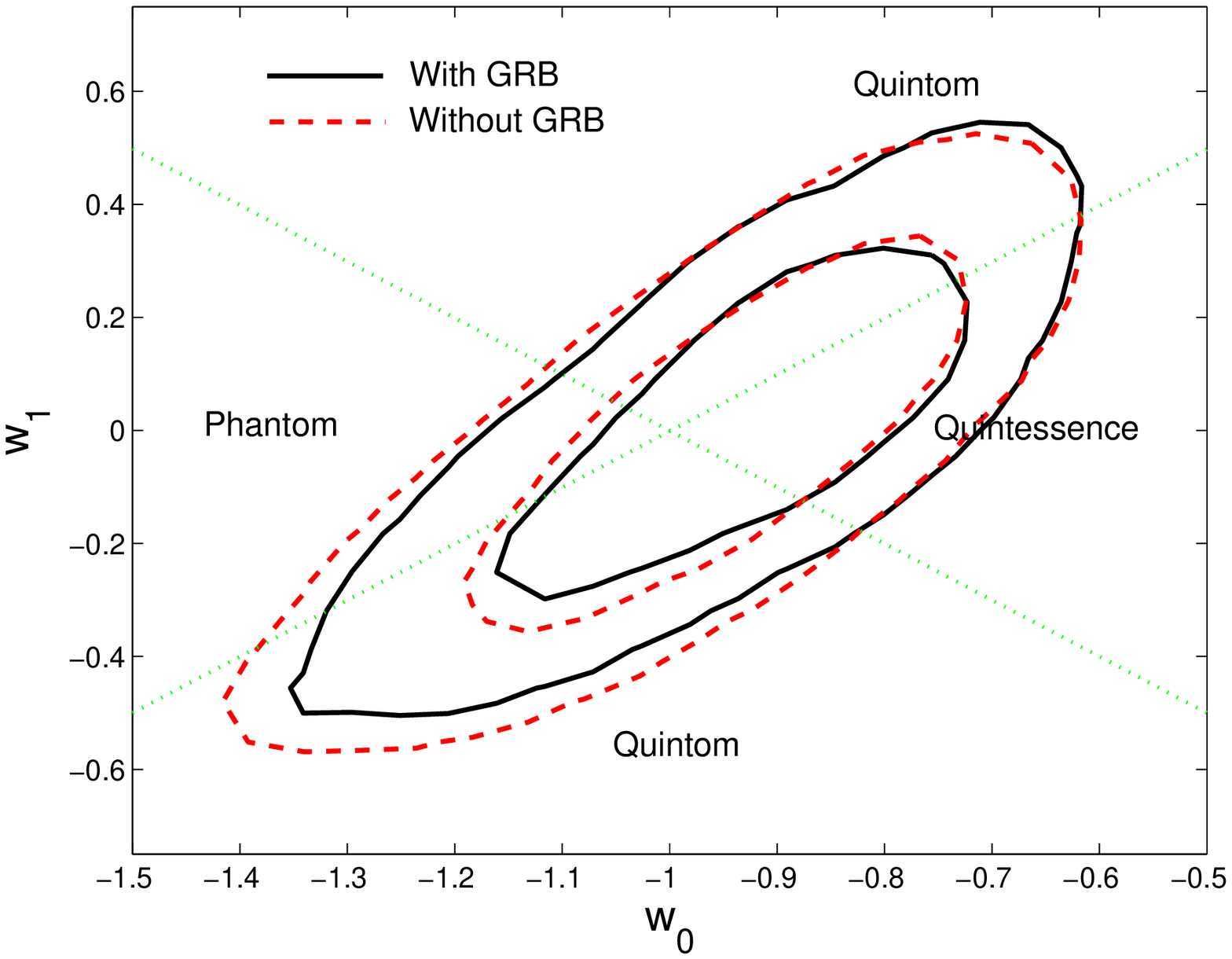}
\caption{2-d joint $68\%$ and $95\%$ confidence regions for the
parameters $w_0$ and $w_1$ of flat model with dark energy Para I
(first panel), non-flat model with dark energy Para I (second
panel), and flat model with dark energy Para II (third panel),
respectively. The black solid line is given by using
WMAP3+SNIa+LSS+GRBs while the red dashed line comes from
WMAP3+SNIa+LSS without GRBs. For both cases we considered the dark
energy perturbation. \label{fig2}}
\end{center}
\end{figure}

It is immediately seen that through constraining the cosmological
parameters, the data sets other than GRB have some effects on the
likelihood of GRB parameters, especially on $C$. Both concerning
the non-flat $\Lambda$CDM model, the black solid line in the right
panel of Figure 1 is much broader than the green solid line. This
is due to the degeneracy between the correlation parameters of GRB
and the cosmological parameters. With GRB data only, cosmological
parameters cannot be tightly constrained, and their large
uncertainties in turn broaden the likelihood of $C$. Including
other data sets greatly reduces the error ranges of cosmological
parameters, resulting a much narrower likelihood distribution for
$C$. It is also noted that the peak of the distribution of $C$
shifts its position as well. This comparison clearly demonstrates
the importance of our methodology in dealing with $A$ and $C$ and
the importance of multi-probe analysis.

On the other hand, given the observational data sets, the
dependence of $A$ and $C$ on cosmological models is rather weak,
as seen from the results shown by the colored lines in Figure 1.
The mean value of $A$ in all four cosmological models is $A\sim
1.53$ with $1\sigma$ error about $0.08$. For the parameter $C$,
its variation for different cosmological models is slightly larger
than that of $A$, with the average value of $C=0.94, 0.96, 0.98$,
and $0.94$, for the four models, respectively. They are in well
agreement with each other within $1\sigma$ range with $\sigma \sim
0.1$. Keeping in mind the hot debate regarding GRBs as
cosmological candles due to the lack of thorough understanding of
GRB physics and the quality of observational data, the consistency
in $A$ and $C$ for different cosmological models seen from our
global fitting analysis may hint that the Ghirlanda relation could
be intrinsic to a subsample of GRBs.

Because we determine $A$ and $C$ simultaneously with the
cosmological parameters, their likelihood distributions shown in
Figure 1 have included the effects of uncertain cosmology
constrained by the current observational data. Therefore our
global fitting analysis give a consistent evaluation on the
contribution of GRBs to the determination of cosmological
parameters. Both the distributions in $A$ and $C$ are broader than
those estimated with a fixed cosmological model. Thus
inappropriate use of the narrow uncertainties in $A$ and $C$
resulting from a given cosmology can lead to an overestimate of
the power of GRBs in cosmological studies.

The results on the cosmological parameters constrained from our
analysis are listed in Table II. In Figure 2, we show,
respectively, the constraints on the dark energy parameters
$w_0-w_1$ for the flat model with dark energy Para I (first
panel), non-flat model with dark energy Para I (second panel), and
flat model with dark energy Para II (third panel), respectively.
The black solid and red dashed lines represent the results with
and without the $27$ GRBs included. For all the cases considered,
the flat $\Lambda$CDM model is well consistent with the
observational data with or without GRBs. For the case of flat
model with Para I for the dark energy EoS, quintom A type of
models with $w_0<-1$ and $w_1>0$ are mildly favored by the data.
Including the GRB data in the analysis, the $1\sigma$ contour
shifts more toward the quintom A region. Relaxing the strong prior
on the flatness of the universe, the $\Lambda$CDM gives a better
fit to the data than the above case. Considering the oscillating
dark energy model, it is seen that the $\Lambda$CDM model with
$w_0=-1$ and $w_1=0$ is again in excellent agreement with the
observational data. Comparing the black solid lines with the red
dashed lines in Figure 2, we see some shrinkage of the error
contours when including GRB data in the analysis, which is largely
attributed to the high redshift range of GRBs. This indicates the
possible potential in using GRBs as high-redshift cosmological
candles. The contribution from the current GRB data is however not
greatly significant. Nevertheless, given the apparent advantage of
GRBs as the tracers of the high-redshift universe, it is important
to perform detailed analysis as we did here to investigate their
usefulness in cosmological studies.

\section{Summary}

In this paper, we present a new method in dealing with the
circular problem for GRBs in the determination of cosmological
parameters. This method is implemented in our MCMC global fitting
program. The methodology is to treat the parameters involved in a
GRB correlation relation as free parameters when performing global
fitting analysis. Their values are then simultaneously estimated
together with the cosmological parameters we are interested in,
and therefore the circular problem is naturally eliminated.
Furthermore, our analysis can give the likelihood distributions of
the correlation parameters with the uncertainties in the
cosmological parameters being taken into account.

From the distributions of $A$ and $C$, we can see that the
dependence of $A$ and $C$ on the cosmology model is rather weak,
and the constraints on $A$ and $C$ for different cosmological
models are in well agreement within $1\sigma$ range. However, the
distributions of $A$ and $C$ are generally broader than the priors
used in many studies in the literature which will lead to the
overestimate of the power of GRBs in cosmological studies. With
the combined datasets CMB+LSS+GRB+SNe, our global fitting results
show that in different dynamical dark energy models the
constraints on dark energy parameters become stringent by taking
into account high redshift GRBs, which show the potential of GRBs
in the cosmology studies.

We emphasize that our method can be readily applied to different
correlation relations of GRBs although we take the Ghirlanda
relation as a concrete example in this paper. In fact, the
applicability of our method is even not limited to GRB studies.
Any cosmological probe involving parameters other than
cosmological ones can be analyzed with our method. Thus our
implemented MCMC program presented in this paper can be a platform
with wide applications. For example, in using the abundance of
clusters of galaxies to constrain cosmology, the relations between
direct observables, such as the X-ray brightness (temperature),
the Sunyaev-Zel'dovich effect and the richness of galaxies, and
the total mass of a cluster have to be involved. Applying the
relations derived based on simplified assumptions regarding the
physical state of clusters may lead to biased cosmological
conclusions. It has been proposed to analyze such relations
simultaneously with cosmological parameters to be studied. Our
program is then perfectly suitable for such analysis.

It is noted that all our investigations and implementations are
carried under the framework of MCMC global fitting using
observational data directly. Therefore we can give more reliable
estimates on the considered parameters than those of Fisher Matrix
analysis or the constraints derived from some extracted
parameters, such as the CMB shift parameter and the BAO parameter
of large-scale structures of the universe.

\acknowledgments

Our MCMC chains were finished in the Sunway system of the Shanghai
Supercomputer Center (SSC). We thank Zi-Gao Dai, Giancarlo
Ghirlanda, Matteo Viel and Dong Xu for helpful discussions. This
work is supported in part by China postdoctoral science
foundation, National Science Foundation of China under Grant Nos.
90303004, 19925523, 10243006, 10373001, 10233010, 10221001 and
10533010, and by the Ministry of Science and Technology of China
under Grant No. NKBRSF G19990754, TG1999075401 and the 973 program
No.2007CB815401, and by the Key Grant Project of Chinese Ministry
of Education (No. 305001). Gong-Bo Zhao is partly supported by the
National Science and Engineering Research Council of Canada
(NSERC).



\begin{thebibliography}{nn}

\bibitem[Bloom et al. (2001)]{Bloom2001}
Bloom, J.~S., Frail, D.~A. \& Sari, R.\ 2001, \aj, 121, 2879

\bibitem[Bloom et al. (2003)]{26}
Bloom, J.~S., Frail, D.~A. \& Kulkarni, S.~R.\ 2003, \apj, 594,
674

\bibitem[Burles et al. (2001)]{BBN}
Burles, S., Nollett, K.~M. \& Turner, M.~S.\ 2001, \apj, 552, L1

\bibitem[Clarkson et al. (2007)]{Clarkson:2007bc}
Clarkson, C., Cortes, M. \& Bassett, B.~A.\ 2007, JCAP, 0708, 011

\bibitem[Cole et al. (2005)]{Cole:2005sx}
Cole, S., et al.\ 2005, \mnras, 362, 505

\bibitem[Dai et al. (2004)]{dailiangxu2004}
Dai, Z.~G., Liang, E.~W. \& Xu, D.\ 2004, \apj, 612, L101

\bibitem[Davis et al. (2007)]{Davis:2007na}
Davis, T.~M., et al.\ 2007, \apj, 666, 716

\bibitem[Di Pietro \& Claeskens (2003)]{DiPietro:2002cz}
Di Pietro, E. \& Claeskens, J.~F.\ 2003, \mnras, 341, 1299

\bibitem[Eisenstein et al. (2005)]{BAO}
Eisenstein, D.~J., et al.\ 2005, \apj, 633, 560

\bibitem[Feng et al. (2005)]{quintom}
Feng, B., Wang, X.~L. \& Zhang, X.\ 2005, \plb, 607, 35

\bibitem[Firmani et al. (2005)]{Firmani2005}
Firmani, C., Ghisellini, G., Ghirlanda, G. \& Avila-Reese, V.\
2005, \mnras, 360, L1

\bibitem[Firmani et al. (2006)]{0610248-36}
Firmani, C., Ghisellini, G., Avila-Reese, V. \& Ghirlanda, G.\
2006, \mnras, 370, 185

\bibitem[Frail et al. (2001)]{Frail:2001qp}
Frail, D.~A., et al.\ 2001, \apj, 562, L55

\bibitem[Freedman et al. (2001)]{Hubble}
Freedman, W.~L., et al.\ 2001, \apj, 553, 47

\bibitem[Friedman \& Bloom (2005)]{friedman2005}
Friedman, A.~S. \& Bloom, J.~S.\ 2005, \apj, 627, 1

\bibitem[Gelman \& Rubin (1992)]{R-1}
Gelman, A. \& Rubin, D.\ 1992, Statistical Science, 7, 457

\bibitem[Ghirlanda et al. (2004a)]{ghirlanda2004a}
Ghirlanda, G., Ghisellini, G., Lazzati, D. \& Firmani, C.\ 2004a,
\apj, 613, L13

\bibitem[Ghirlanda et al. (2004b)]{0610248-34}
Ghirlanda, G., Ghisellini, G. \& Lazzati, D.\ 2004b, \apj, 616,
331

\bibitem[Ghirlanda et al. (2007)]{ghirlanda0702352}
Ghirlanda, G., Nava, L., Ghisellini, G. \& Firmani, C.\ 2007,
A\&A, 466, 127

\bibitem[Goliath et al. (2001)]{goliath2001}
Goliath, M., Amanullah, R., Astier, P., Goobar, A. \& Pain, R.\
2001, A\&A, 380, 6

\bibitem[Huterer \& Cooray (2005)]{Huterer:2004ch}
Huterer, D. \& Cooray, A.\ 2005, \prd, 71, 023506

\bibitem[Lewis \& Bridle (2002)]{CosmoMC}
Lewis, A. \& Bridle, S.\ 2002, \prd, 66, 103511; See also the
CosmoMC website at: http://cosmologist.info.

\bibitem[Li et al. (2008)]{grb0612060}
Li, H., Su, M., Fan, Z.~H., Dai, Z.~G. \& Zhang, X.\ 2008, \plb,
658, 95

\bibitem[Lloyd-Ronning \& Ramirez-Ruiz (2002)]{0610248-96}
Lloyd-Ronning, N.~M. \& Ramirez-Ruiz, E.\ 2002, \apj, 576, 101

\bibitem[Miknaitis et al. (2007)]{Miknaitis:2007jd}
Miknaitis, G., et al.\ 2007, \apj, 666, 674

\bibitem[Norris et al. (2000)]{0610248-73}
Norris, J.~P., Marani, G.~F. \& Bonnell, J.~T.\ 2000, \apj, 534,
248

\bibitem[Riess et al. (2007)]{Riess:2006}
Riess, A.~G., et al.\ 2007, \apj, 659, 98

\bibitem[Schaefer (2007)]{schaefer2006}
Schaefer, B.~E.\ 2007, \apj, 660, 16

\bibitem[Spergel et al. (2007)]{Spergel:2006hy}
Spergel, D.~N., et al.\ 2007, \apjs, 170, 377

\bibitem[Su et al. (2006)]{sumeng0611155}
Su, M., Li, H., Fan, Z.~H. \& Liu, B.\ 2006,
arXiv:astro-ph/0611155

\bibitem[Tegmark et al. (2004)]{Tegmark:2003ud}
Tegmark, M., et al.\ 2004, \prd, 69, 103501

\bibitem[Tegmark et al. (2006)]{Tegmark:2006az}
Tegmark, M., et al.\ 2006, \prd, 74, 123507

\bibitem[Wang et al. (2007)]{wangfayin2007}
Wang, F.~Y., Dai, Z.~G. \& Zhu, Z.~H.\ 2007, \apj, 667, 1

\bibitem[Wang \& Mukherjee (2006)]{shift_R}
Wang, Y. \& Mukherjee, P.\ 2006, \apj, 650, 1

\bibitem[Wright (2007)]{wright}
Wright, E.\ 2007, \apj, 664, 633

\bibitem[Xia et al. (2006)]{Xia:2005ge}
Xia, J.~Q., Zhao, G.~B., Feng, B., Li, H. \& Zhang, X.\ 2006,
\prd, 73, 063521

\bibitem[Xu et al. (2005)]{xudong2005}
Xu, D., Dai, Z.~G. \& Liang, E.~W.\ 2005, \apj, 633, 603

\bibitem[Zhang et al. (2006)]{Zhang:2005eg}
Zhang, X.~F., Li, H., Piao, Y.~S. \& Zhang, X.\ 2006, \mpla, 21,
231

\bibitem[Zhao et al. (2005)]{Zhao:2005vj}
Zhao, G.~B., Xia, J.~Q., Li, M., Feng, B. \& Zhang, X.\ 2005,
\prd, 72, 123515

\bibitem[Zhao et al. (2007a)]{Zhao:2007a}
Zhao, G.~B., et al.\ 2007a, \ijmpd, 16, 1229

\bibitem[Zhao et al. (2007b)]{Zhao:2007b}
Zhao, G.~B., et al.\ 2007b, \plb, 648, 8

\end{thebibliography}
\end{document}